\documentclass[aps,prb,twocolumn,showpacs,superscriptaddress,groupedaddress]{revtex4-1}  
\usepackage{graphicx}  
\usepackage{dcolumn}   
\usepackage{bm}        
\usepackage{amsmath}
\usepackage{amssymb}   
\usepackage{feynmp}
\usepackage{paralist, caption, subcaption, placeins}
\renewcommand{\vec}[1]{\mathbf{#1}}
\newcommand{\gvec}[1]{\boldsymbol{#1}}

\newcommand{\lambdav}{\boldsymbol\lambda}

\newcommand{\Ut}{\tilde{U}}
\newcommand{\Deltat}{\tilde{\Delta}}

\newcommand{\epsdts}{\tilde{\epsilon}_{d, \sigma}}
\newcommand{\epsdt}{\tilde{\epsilon}_d}

\newcommand{\Sigmat}{\tilde{\Sigma}}
\newcommand{\Gammat}{\tilde{\Gamma}}
\newcommand{\Ueff}{\tilde{U_e}}
\newcommand{\Pit}{\tilde{\Pi}}
\newcommand{\rmd}{\textrm{d}}
\newcommand{\GeeNt}{\tilde{G}^{[0]}}

\newcommand{\sgn}{\operatorname{sign}}
\newcommand{\Arg}{\operatorname{Arg}}

\renewcommand{\Im}{\textrm{Im}}
\renewcommand{\Re}{\textrm{Re}}

\begin{document}
\widetext
\leftline{Current as of \today}

\title{High-order terms in the renormalized perturbation theory for the Anderson impurity model}

\author{Vassilis Pandis, Alex C.\ Hewson}
\affiliation{Department of Mathematics, Imperial College London, London SW7 2AZ, United Kingdom}
\date{\today}


\begin{abstract}
We study the renormalized perturbation theory of the single-impurity Anderson model, particularly the high-order
terms in the expansion of the self-energy in powers of the renormalized coupling $\Ut$. 
Though the 
presence of counter-terms in the renormalized theory may appear to complicate the diagrammatics, we show how these can be seamlessly
accommodated by carrying out the calculation order-by-order in terms of skeleton diagrams. We describe how the diagrams pertinent to the renormalized self-energy and four-vertex
can be automatically generated, translated into integrals and numerically integrated. To maximize the efficiency of our approach
we introduce a generalized $k$-particle/hole propagator, which is used to analytically simplify the resultant
integrals and reduce the dimensionality of the integration. We present results for the self-energy and spectral
density to fifth order in $\Ut$, for various values of the model asymmetry, and compare them to a Numerical Renormalization Group
calculation. 
\end{abstract}

\maketitle

\section{Introduction}\label{sec:intro}

The Anderson Impurity Model (AIM)~\cite{Anderson61} was introduced in $1961$ to explain
the existence of localized magnetic moments in metals with magnetic impurities. It stands today as 
one of the best understood impurity models, having been intensely studied by a number of methods such as the bare
perturbation theory~\cite{Yosida70, Yamada75, Yosida75, Horvatic80, Horvatic82},  $1/N$ expansions~\cite{Bickers87} (where $N$ denotes the impurity
orbital degeneracy), the Bethe Ansatz~\cite{Andrei83, Wiegmann83} and the Numerical Renormalization Group (NRG) ~\cite{Wilson75, Krishna-murthy80} (among many others; see Ref.~\cite{Hewson97} for a review). 
Each of these is particularly useful in different scenarios: the bare perturbation theory excels at weak-coupling, $1/N$ expansions are asymptotically
correct for highly degenerate models, the Bethe Ansatz yields exact results, but only for static quantities, and the
NRG approach cannot efficiently deal with large degeneracies. Together, these methods paint a consistent picture for the physics of the model, but their
abundance highlights the challenges posed even by relatively simple models of strongly correlated electrons. This in-depth
understanding of the model has served to establish it as a testing ground for the development of new methods in many-body physics.   
Though the AIM is today well understood in the context of a single impurity, interest in it has been renewed in light of the
Dynamical Mean-Field Theory~\cite{Georges96}, which allows models of impurity lattices to be mapped --- exactly in the limit of infinite dimensions --- onto a single-impurity 
problem subject to a self-consistency consistency condition. 

An important step in the understanding of impurity physics was the connection made by Nozi\`{e}res between the strong coupling fixed point of the Kondo model at low temperatures and Landau's theory
of Fermi liquids~\cite{Nozieres74}, in which the excited states of the model are interpreted in terms of quasi-particles
which are weakly interacting\footnote{This follows from a phase-space argument and does not imply that the interaction constant is small.}, even in the strong correlation regime. It was quickly recognized that the AIM is a Fermi 
Liquid in all parameter regimes~\cite{Haldane78, Krishna-murthy80} though this idea was not fully explored until very recently~\cite{Sela09, Mora14}.

In light of the Fermi liquid interpretation, the Renormalized Perturbation Theory (RPT)~\cite{Hewson93b, Hewson94, Hewson01, Hewson04, Hewson06, Hewson06b,Hewson06c,  Edwards11b, Edwards13} 
offers a convenient way of
analysing the low-energy behaviour of the model. The usual Hamiltonian $H$ of the AIM specifies the model in 
terms of the impurity level $\epsilon_d$, the hybridization broadening $\Delta$ and the local Coulomb interaction $U$. 
In the renormalized theory these are replaced by effective parameters $\epsdts, \Deltat_\sigma, \Ut$ which are used to define
a renormalized Hamiltonian $\tilde{H}$ of the same form as $H$ and which, by definition, incorporates the low-energy
one-particle interactions. The propagators in the renormalized theory describe the quasi-particle
states, making explicit the one-to-one correspondence between the single-particle excitations of the non-interacting 
and interacting systems. The quasi-particle interactions can now be taken into account by constructing a perturbation
theory in the renormalized parameters and organized in powers of $\Ut$.  

The RPT has a number of appealing features~\cite{Hewson93b}. In the presence of a magnetic field the leading term of the 
renormalized expansion is of order $\Ut$, and suffices to calculate the zero-temperature spin and charge susceptibilities exactly in all 
parameter regimes. In the absence of a magnetic field the leading term in the RPT is of order $\Ut^2$ and leads to a simple exact expression for the second
derivative of the imaginary part of the self-energy at zero frequency. This in turn can be used to derive an exact expression for the $T^2$ 
coefficient of the conductivity of the symmetric  model in terms of the renormalized parameters. 

In this paper we discuss the calculation of the renormalized self-energy using the diagrammatic RPT and show how this can be implemented
on a computer completely automatically.  Our presentation is structured as follows: We start by describing a simple algorithm to generate all relevant Feynman diagrams
and retain only those that contribute to the renormalized self-energy. Though at first the RPT seems to have a more complicated perturbational 
structure than the bare theory, owing to the presence of counter-terms, we show how these can be included into the calculation with minimal effort by setting up
the calculation in terms of skeleton diagrams. As an intermediate step we introduce a simplification algorithm which dramatically reduces the computational 
complexity of the resultant integrals by factorising out sub-integrations that can be computed analytically. Finally, we carry out the numerical 
integrations and present results to order \emph{$\Ut^5$} inclusive for the self-energy and spectral density in the strong correlation regime, for different 
values of the asymmetry, and compare these to results obtained using the NRG.

\section{Renormalized Perturbation Theory}

The effective Lagrangian for the Anderson model in the limit of an infinitely wide conduction electron band is~\cite{Hewson01}
\begin{align}
\mathcal{L}  = \sum_{\sigma = \uparrow, \downarrow} \overline{d}_\sigma(\tau) & \left(  \partial_\tau - \epsilon_{d, \sigma} + i\Delta\right) d_\sigma(\tau) \nonumber \\
&+ U n_{\uparrow}(\tau) n_{\downarrow}(\tau).
\label{eq:lagbare}
\end{align}
In the bare perturbation theory one attempts to solve Eq.~\eqref{eq:lagbare} by taking
the $U=0$ (or Hartree-Fock) state as the starting point for a perturbation theory in powers of the interaction $U$. Though 
some useful information can be extracted this way in the weak correlation regime, the value of $U$ in physically relevant systems is usually too large
to be handled in this manner; this is precisely the challenge of strongly correlated physics. In particular, it has long 
been recognized that the low-energy excitations primarily responsible for the interesting physics of impurity models
cannot be described perturbatively. 

The difficulty of the bare perturbation theory can be ultimately traced to the unfortunate choice of the $U=0$ state as the starting point for the
perturbation expansion. As we increase $U$, virtual low-energy scattering processes will lead to the formation 
of quasi-particles on the impurity site which interact through a renormalized Coulomb interaction $\Ut$,  and whose relation to the original
particles becomes increasingly tenuous. The RPT seeks to address this issue by 
using precisely these quasi-particle states as the starting point for the perturbation expansion in powers of the renormalized coupling. This is accomplished by
writing the Lagrangian in Eq.~\eqref{eq:lagbare} as a sum of a renormalized quasi-particle Lagrangian, which describes the quasi-particles, a
renormalized interaction term and a remainder term, the \emph{counter-term} Lagrangian, i.e.\ 
$\mathcal{L} = \tilde{\mathcal{L}_0} +  \tilde{\mathcal{L}_U} + \mathcal{L}_{\textrm{ct}}$, where 
\begin{align}
\tilde{\mathcal{L}}_0 &=  \sum_{\sigma = \uparrow, \downarrow} \tilde{\overline{d}}_\sigma(\tau) \left(\partial_\tau - \epsdts + i\Deltat_\sigma\right) d_\sigma(\tau), \label{eq:l0} \\
\tilde{\mathcal{L}}_{ct} &=  \sum_{\sigma = \uparrow, \downarrow}\tilde{\overline{d}}_\sigma(\tau) (\lambda_{2, \sigma} \partial_\tau + \lambda_{1, \sigma})\tilde{d}_\sigma(\tau) + \lambda_3 \tilde{n}_\uparrow(\tau) \tilde{n}_\downarrow(\tau), \\
\tilde{\mathcal{L}}_U &= \Ut \tilde{n}_\uparrow \tilde{n}_\downarrow.
\end{align}
In this re-arrangement the term $\tilde{\mathcal{L}}_0$ forms the starting point for the perturbative expansion and the interaction term is taken to be $\tilde{\mathcal{L}_U} + \tilde{\mathcal{L}}_{\textrm{ct}}$. This
term gives rise to a renormalized self-energy $\Sigmat_\sigma(\omega)$ and two-particle-reducible four-vertex $\Gammat_{\uparrow \downarrow}(\omega_1, \omega_2; \omega_3, \omega_4)$. 
The counter-terms $\lambdav = (\lambda_{1, \sigma}, \lambda_{2, \sigma}, \lambda_3)$ are then determined by imposing the renormalization conditions
\begin{align}
\Sigmat_\sigma (0) &= 0, \nonumber \\
\partial_\omega\Sigmat_\sigma(\omega)|_{\omega=0} &= 0, \nonumber \\
\Gammat_{\uparrow \downarrow}(0,0;0,0)  &= \Ut.
\label{eq:RGeq}
\end{align}
The presence of the counter-terms ensures that renormalization effects that have already been absorbed in the renormalized parameters are not overcounted. 
The counter-terms are real numbers; this is due to the Fermi liquid property of the AIM.

In many ways, this re-organization of the Lagrangian in terms of the renormalized parameters is similar to the corresponding practice
in High Energy Physics (see Ref.~\cite{Peskin95} for instance). The motivation is largely the same, to re-express the Lagrangian in terms of physical parameters. Nevertheless
there are conspicuous differences. For instance, in our system there are natural upper and lower energy cut-offs provided by the metal's lattice spacing
and sample size respectively. Therefore the technicalities of the regularization procedure, which can be rather complex for many-loop calculations, 
do not enter our discussion at all. 

For the purposes of this article we use the NRG to determine the renormalized parameters~\cite{Hewson04}; note however, that
in certain cases they can be determined entirely within RPT without appealing to an external method~\cite{Edwards11b, Edwards13, Pandis14}. 
Given the renormalized parameters, Eq.~\eqref{eq:RGeq} is imposed in some approximation for $\Sigmat_\sigma(\omega)$; the counter-terms thus depend on the
chosen approximation to the self-energy, whereas the renormalized parameters are independent of it, and are in a one-to-one correspondence with the
bare parameters that define the model. We can relate the renormalized parameters $\epsdts, \Deltat_\sigma$ to the bare ones through
\begin{align}\label{eq:renormdef}
\epsdts &= z_\sigma\left(\epsilon_{d,\sigma} + \Sigma_\sigma(0)\right), \nonumber  \\
\Deltat_{\sigma} &= z_\sigma \Delta,
\end{align}
where 
\begin{equation}
z_\sigma = \frac{1}{1 - \partial_\omega\Sigma_\sigma(\omega)|_{\omega=0}},
\end{equation}
and similarly the relate the renormalized self-energy to the bare quantity through
\begin{equation}
\Sigmat_\sigma(\omega) = z_\sigma \left(\Sigma_\sigma(\omega) - \omega \partial_\omega \Sigma_\sigma(\omega)|_{\omega=0} - \Sigma_\sigma(0)\right).
\label{eq:sigmatdef}
\end{equation}
We eliminate the quantities $\partial_\omega\Sigma_\sigma(\omega)|_{\omega=0}$ and $ \Sigma_\sigma(0)$ in favour of $\epsdt$ and $ z_\sigma$ to obtain
\begin{equation}
\epsdts + \Sigmat_\sigma(\omega) = z_\sigma\left(\Sigma_\sigma(\omega) + \epsilon_{d, \sigma}\right)  - (1-z_\sigma)\omega.
\label{eq:sigmarelation}
\end{equation}
From Eqs.~\eqref{eq:renormdef},~\eqref{eq:sigmatdef} we find the renormalized interacting propagator
\begin{equation}
\tilde{G}_\sigma(\omega) = [\omega - \epsdts + i\Deltat_\sigma- \Sigmat_\sigma(\omega)]^{-1} = z^{-1}_\sigma G_\sigma(\omega) 
\label{eq:grpt}
\end{equation}
and thus deduce via a Fourier transform and the definition of the Green's function $G_\sigma(\tau) = \langle \tilde{d}_\sigma(\tau)\tilde{\overline{d}}_\sigma(0)\rangle$ that 
$\tilde{d}_\sigma(\tau) = z_\sigma^{1/2}d_\sigma(\tau)$ and $\tilde{\overline{d}}_\sigma(\tau) = z_\sigma^{1/2}\overline{d}_\sigma(\tau)$.

\section{Automated RPT expansions}

In this section we will describe the automation of the calculation of the self-energy. For the purposes of our calculation we will use the $T=0$ formalism. The renormalized Green's function which will form the basis of the diagrammatic expansion is 
\begin{equation}
\GeeNt_\sigma(\omega) = \frac{1}{\omega - \epsdts + i \Deltat_\sigma\sgn(\omega)}.
\label{eq:greens}
\end{equation}
The $T=0$ formalism has the advantage of working directly on the real axis, thus avoiding the need for an analytic continuation, 
which is often fraught with its own numerical difficulties. More generally, for $m\geq1$ we introduce
the propagator
\begin{equation}
\tilde{G}^{[m]}_\sigma(\omega) = \GeeNt_\sigma(\omega) \Sigmat_\sigma^{[m]}(\omega) \GeeNt_\sigma(\omega),
\end{equation}
where $\Sigmat_\sigma^{[m]}(\omega)$ denotes the $\Ut^m$ (strictly) term of the renormalized self-energy. 
To simplify the discussion we assume a zero magnetic field, so $\epsdts$ and $\Deltat_\sigma$ are spin-independent quantities.

Our diagrammatic approach is based on the skeleton formalism. This has the advantage of avoiding explicit
reference to the counter-terms $\lambda_1$ and $\lambda_2$. Furthermore, we introduce an effective 
interaction constant $\Ueff = \Ut + \lambda_3$ that combines the $\lambda_3$ counter-term with the 
renormalized interaction constant. This allows us to tentatively carry out the expansion of $\Sigma(\omega)$
in powers of $\Ueff$, without having to explicitly account for $\lambda_3$. Ultimately, our goal is to organize 
the calculation in powers of $\Ut$, rather than $\Ueff$. This will be discussed at the end of this section, but we 
note for now that it requires knowledge of $\Gammat_{\sigma, -\sigma}(0,0;0,0)$ order-by-order in $\Ueff$, up to 
$\Ueff^4$ inclusive. We thus aim to generate and calculate diagrams for the self-energy and the four-vertex.

\subsection{Generating the diagrams}

\begin{figure}
\begin{center}
\includegraphics[scale=0.25]{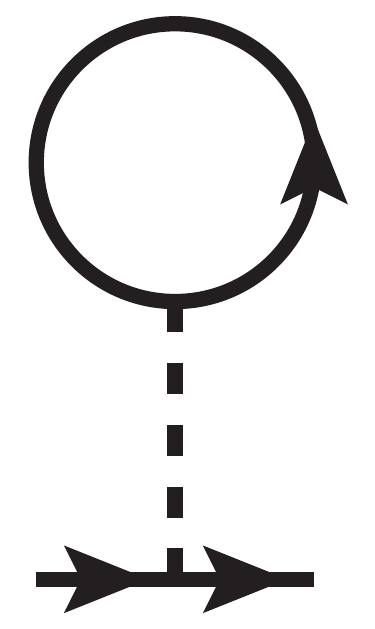}
\end{center}
\caption{The first-order correction to the renormalized self-energy. The internal line corresponds to the free quasi-particle
propagator of Eq.~\eqref{eq:greens} and the broken line denotes the effective interaction vertex $\Ueff$.}
\label{fig:loop}
\end{figure}

Consider the first-order term given by the loop diagram of Fig.~\ref{fig:loop} and a tree-level
counter-term. Since the loop is frequency independent, we see, upon imposing Eq.~\eqref{eq:RGeq},
that the first-order self-energy is $\Sigmat_\sigma^{[1]}(\omega) = 0$. Since $\Gammat_{\uparrow, \downarrow}(0,0;0,0)$ is by definition equal to $\Ut$, we find 
that $\lambda_3 = \mathcal{O}(\Ut^2)$. Note that the cancellation of the loop to this order implies that
\emph{any} diagram with a loop will cancel with the corresponding diagram that has a counter-term in place of the loop, \emph{to all orders} in $\Ut$.
For similar reasons we can ignore diagrams whose external legs attach to the same interaction vertex, for
these do not depend on the external frequency and will cancel with the counter-term. In this sense, calculations in the renormalized
theory thus involve fewer diagrams than the bare theory. 

To calculate the self-energy as a power series in $\Ut$ we have, in principle, to explicitly include
the six interaction vertices --- five from the possibly spin-dependent counter-terms  and the sixth being $\Ut$ --- order-by-order in the
renormalized expansion. Fortunately, in the absence of a magnetic field, explicit reference to the counter-term vertices can be avoided by setting up the
perturbation theory in the skeleton formalism. We define a \emph{skeleton diagram} as a diagram that does not contain self-energy insertions. A skeleton diagram
involving $n$ interaction vertices will necessarily contain $2n-1$ internal lines. By replacing all of them with
$\GeeNt$ we construct a self-energy diagram of order $n$; by replacing any $2n-2$ lines with $\GeeNt$ and the
remaining line with $\tilde{G}^{[1]}$ we obtain one of the diagrams that contribute to $\Sigmat^{[n+1]}$, and so forth.
An advantage of the skeleton formalism is that it minimizes the number of diagrams that have to
be accounted for, since most of the diagrams of order $n$ can be generated from a lower-order skeleton diagram. 
Note that for $n<6$ we do not need to consider diagrams with more than one insertion, since a second-order
diagram with two second-order insertions would contribute a $\Ut^6$ term.

Rather than attempt to enumerate all relevant diagrams manually,
which is tedious and error prone, we used the \textsc{igraph} library~\cite{igraph} to first construct all possible diagrams and then retain only those
relevant to our calculation. Since we are only interested in relatively small orders $n$ --- the bottleneck is the numerical integration, not
the diagram generation --- we found it sufficient
to generate diagrams by simply exhaustively considering all possible combinations of connecting $n$ vertices such that
no internal line begins and ends on the same vertex. By regarding each Feynman diagram as a directed graph with multiple edges,
and calculating the graph's edge connectivity, we can identify and discard one-particle reducible diagrams.

In graph-theoretic terms a skeleton graph can be defined as a  graph that does not
contain a subgraph isomorphic to a lower-order skeleton graph. Consequently, to identify the
skeleton graphs of order $n$ we must generate and store all diagrams of order~\footnote{Diagrams
that contain insertions of order $n-1$ can only produce a static contribution.} $2,3,\ldots, n-2$. To identify the non-skeleton diagrams we then
bijectively map each (multi-edged) Feynman graph to a simple graph with weights encoding the spins of the multiple edges and
apply the \textsc{VF2} algorithm~\cite{VF2A, VF2B} for the subgraph isomorphism problem to the simple graphs (this `edge coloring' technique
is necessary to apply the algorithm to multi-edge graphs). Working
order-by-order we can thus generate all the skeleton diagrams up to our desired order.  

The diagrams that contribute to the two-particle-reducible four-vertex can be generated in a similar way.
Whereas in the case of the self-energy we only had to consider the possibility that 
two external legs attach to different interaction vertices, the diagrams of the four-vertex admit
more complex topologies which have to be taken into account individually. The calculation is otherwise
identical, except for the fact that there are only $2n-2$ internal lines corresponding to a diagram 
of order $n$. Our method can similarly be extended to higher correlation functions, but this is beyond the scope of this paper.  
In Table~\ref{table:numsigma} we report the number of relevant diagrams to fifth order in $\Ueff$.
\begin{table}
\centering
\begin{tabular}{ l |  c | r } 
V & $\Sigmat$ & $\Gammat$ \\
\hline
2 &  1  & 2 \\
3 & 2   &  9 \\ 
4 & 12  &  58 \\ 
5 & 73  & 438 \\
\end{tabular} 
\caption{Number of dynamic skeleton self-energy and four-vertex diagrams as a function of the number of vertices $V$.}
\label{table:numsigma}
\end{table}

\subsection{Evaluating the diagrams}

Having generated all relevant self-energy diagrams we proceed to impose frequency conservation at
each vertex. An $n$'th order diagram for the self-energy will have $2n-1$ internal lines. These are subject
to $n$ constraints, though due to global frequency conservation only $n-1$ are independent.
We thus have $2n-1 - (n-1)= n$ independent frequencies, each of which corresponds to an integration
variable.   We thus arrive at a linear, underdetermined,  system of dependent equations, $\gvec{c}\gvec{\omega} = \gvec{\Omega}$, where
$\gvec{c}$ is an $n\times(2n-1)$ matrix encoding the constraints in the form $\sum(\textrm{in}) - \sum({\textrm{out})}$, $\gvec{\omega}$ a vector with
$2n-1$ components, each corresponding to the frequency of an internal line,  and $\gvec{\Omega}$ the $n$-dimensional vector $(-\Omega, 0, 0, \ldots, \Omega)$, where
$\Omega$ is the external frequency~\footnote{This form of $\gvec{\Omega}$ follows by adopting the convention to place the constraints emerging from frequency conservation on the 
vertices that attach to the incoming and outgoing legs on the first and last row of $\vec{c}$ respectively.}. This can now be readily inverted to give the internal line
frequencies as a function of the $n$ integration variables $\gvec{\epsilon}$  and a particular
solution $\gvec{\Omega}_p$ to the system\footnote{Note that the matrix $\gvec{f}$ is not unique but
we can always perform a unitary transformation so that all its entries are equal to $0$ or $\pm 1$.}
\begin{equation}
\gvec{\omega} = \gvec{f}\gvec{\epsilon} +  \gvec{\Omega}^{(p)}.
\label{eq:fmatrix}
\end{equation} 
This can be accomplished in a number of ways; we found it best to determine the null space using the Lenstra-Lenstra-Lov\'{a}sz lattice
basis reduction algorithm~\cite{LLL}, as this results in $\gvec{f}$-matrices similar to what one would obtain by manually
imposing frequency conservation (i.e.\ matrices whose entries are $\pm1$ or $0$). 
The diagrams for the four-vertex are treated similarly: an $n$'th order diagram 
will involve $n-1$ integration variables, resulting in an $n \times (2n-2)$ $\gvec{c}$-matrix and a $(2n-2) \times (n-1)$ $\gvec{f}$-matrix.

It is possible, though very inefficient, to apply the Feynman rules at this point
and proceed with the numerical integration. In calculations by hand, however, it is common 
to analytically factorize out any particle-hole or particle-particle pair-propagators.
This has the advantage of reducing the dimension of the
numerical integration and consequently reducing the number of integrand evaluations
needed to achieve a given precision. We can generalize this to accommodate the more general
scenario of $k$ (quasi-)particle/hole lines and define
\begin{equation}
\Pit^{(k)}_{\vec{\sigma}; \vec{s}}(\omega_1, \ldots, \omega_k) = i^k\int_{-\infty}^\infty \rmd  \omega' \prod_{i=1}^{k} G^{[0]}_{\sigma_i} (s_i\omega' + \omega_i),
\label{eq:npp}
\end{equation}
where $s_i = 1$ for particle lines and $s_i=-1$ for hole lines. Due to the simple form of the propagator, 
the integral can be calculated analytically; the details of the calculation have been relegated to the Appendix. 

Having defined the $k$-particle/hole propagator we will now describe an algorithm to identify instances of it from the information 
encoded in the $\gvec{f}$-matrix. Our strategy will be to first inspect the $\gvec{f}$-matrix and identify groups of Green's functions
that can be combined to form a product of the form of Eq.~\eqref{eq:npp}. We accomplish this by traversing the $\gvec{f}$-matrix column-by-column 
--- since each column corresponds to an integration variable --- and examining each column's non-zero entries. Our goal is ultimately to delete
from the $\gvec{f}$-matrix the columns which correspond to integration variables that have been absorbed in $\Pit^{(k)}_{\vec{\sigma}; \vec{s}}$,
and to delete the rows that correspond to Green's functions that comprise $\Pit^{(k)}_{\vec{\sigma}; \vec{s}}$.
To avoid modifying the $\gvec{f}$-matrix  while we are traversing it we will maintain a list $L_c$ of columns  and a list $L_r$ of rows which are 
to be removed, both of which are empty at the start of the search, and we will only delete the rows and columns after all possible
simplifications have been identified. We will refer to the resultant simplified matrix as the $\vec{g}$-matrix, to distinguish it from the 
original $\gvec{f}$-matrix.

To identify instances of $\Pit^{(k)}_{\vec{\sigma}; \vec{s}}$ we scan each column $m$ of the $\gvec{f}$-matrix and identify
the columns with exactly $k$ non-zero entries $\mu_1, \ldots, \mu_k$ on the rows $r_1, \ldots, r_k$ 
\footnote{Since we are  using the LLL algorithm to find $\gvec{f}$ we also know that $|\mu_1| = \ldots = |\mu_k| = 1$, since any non-zero
entry will be $\pm1$.}. If $m \not\in L_c$ and $r_1, \ldots, r_k \not\in L_R$ then column $m$ indeed corresponds to a $k$-particle/hole propagator;
if not, and either $m \in L_c$ or at least one of $r_1, \ldots r_k \in L_R$, it means that the corresponding Green's function has already been absorbed into another propagator, and we proceed to the next column.
With the $j$'th $k$-particle/hole propagator\footnote{The variable $j$ here is merely a label; the first $k$-particle/hole propagator we identify is $j=1$,  and so on.} 
we will associate a matrix $\vec{q}_j$,  which encodes its dependence on the remaining integration 
variables, and vectors $\vec{\sigma}_j$, $\vec{s}_j$ describing the spin and sign configuration respectively. 
The $k$ frequency arguments of $\Pit^{(k_j)}_{\vec{\sigma}_j; \vec{s}_j}$ will therefore be given by the rows of $\vec{q}_j$.
Furthermore, we construct a $k$-dimensional vector
$\gvec{\Omega}_\vec{q}$ that contains the external frequency dependence of the rows $r_1, \ldots, r_k$, i.e.\ $\gvec{\Omega}_\vec{q} = (\Omega^{(p)}_{ r_1}, \ldots, \Omega^{(p)}_{r_k})$. 
In other words, the $\omega_i$ that appear in Eq.~\eqref{eq:npp} can be obtained from the $i$'th component of 
\begin{equation}
\vec{q}\gvec{\epsilon}' + \gvec{\Omega}_\vec{q},
\label{eq:qeq}
\end{equation}
where $\gvec{\epsilon}'$ denotes
the free variables that remain after all propagator simplifications.  After constructing the $\vec{q}-$matrix we add $m$ to $L_c$ and
all of $r_1, \ldots, r_k$ to $L_r$ and examine the next column. When the columns of the matrix have been exhausted we
delete all columns in $L_c$ and rows in $L_r$ from the original $\gvec{f}$-matrix, to obtain its final form $\vec{g}$. Furthermore, to
account for the fact that components of the original $\gvec{\Omega}^{(p)}$ have been absorbed into the various $\gvec{\Omega}_\vec{q}$ we 
delete all the rows in $L_r$ from $\gvec{\Omega}^{(p)}$ to obtain its final form, which we denote $\gvec{\Omega}^{(g)}$. Finally, we eliminate
the columns in $L_c$ from the provisional $\vec{q}$ matrices. 

For each diagram we thus arrive at an expression for the amplitude of the form
\begin{equation}
I_\mathcal{D} = P_\mathcal{D} \int \rmd\gvec{\epsilon} \tilde{G}^{[l_1]}_{\sigma_1} \tilde{G}^{[l_2]}_{\sigma_2} \ldots \Pit^{(k_1)}_{\vec{\sigma}_1; \vec{s}_1} \Pit^{(k_2)}_{\vec{\sigma}_2;\vec{s}_2} \ldots , 
\label{eq:integrand1}
\end{equation}
where $P_\mathcal{D}$ is a complex prefactor containing powers of $\Ueff, i$ and $2\pi$. In Eq.~\eqref{eq:integrand1} it is understood that the argument
of $\tilde{G}^{[l_i]}_{\sigma_i}$ corresponds to the $i$'th row of 
\begin{equation}
\gvec{\omega} = \gvec{g}\gvec{\epsilon} +  \gvec{\Omega}^{(g)},
\end{equation}
which is similar to Eq.~\eqref{eq:fmatrix} but involves $\vec{g}$, the simplified form of the $\gvec{f}$-matrix, and that the (vector) argument
of the $j$'th $\Pit^{(k_j)}_{\vec{\sigma_j}_; \vec{s}_j} $ is given by Eq.~\eqref{eq:qeq}, with the corresponding $\gvec{q}_j$ and $\gvec{\Omega}_{\vec{q}_j}$. 
We remark that while the final integral of Eq.~\eqref{eq:integrand1} 
is at least one-dimensional for all self-energy diagrams, some four-vertex diagrams factorize completely.

\begin{figure}
\begin{center}
\includegraphics[scale=0.25]{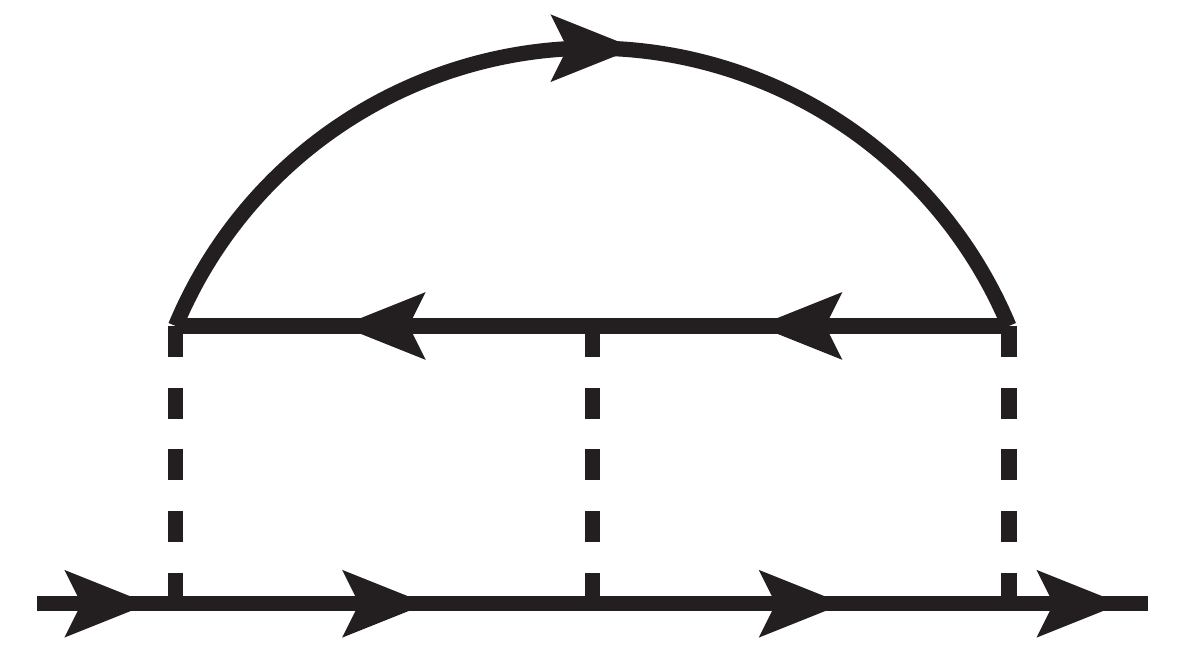}
\end{center}
\caption{One of the two diagrams that contribute to the self-energy to third order in $\Ueff$.}
\label{fig:examplediagram}
\end{figure}

To illustrate this with an example consider the diagram of Fig.~\ref{fig:examplediagram}. 
We find that the amplitude is proportional to 
\begin{align}
\int \rmd\gvec{\epsilon}\GeeNt_\sigma(\epsilon_1 - \epsilon_3 + &\Omega) \GeeNt_\sigma(\epsilon_2 - \epsilon_3 + \Omega) \nonumber \\
                &\times \GeeNt_{-\sigma}(\epsilon_1)\GeeNt_{-\sigma}(\epsilon_2)\GeeNt_{-\sigma}(\epsilon_3),
\label{eq:exampleeq}
\end{align}
or in matrix notation,
\begin{equation}
\gvec{f} = \begin{bmatrix}
1 & 0 & -1 \\
0 & 1 & -1 \\
1 & 0 &  0 \\
0 & 1 & 0  \\
0 & 0 & 1 \\
\end{bmatrix}, 
\quad
\gvec{\Omega}^{(p)} = \begin{bmatrix}
\Omega \\ \Omega \\ 0 \\ 0 \\ 0 
\end{bmatrix}.
\label{eq:fmatrixeq}
\end{equation} 
We begin with the first column and identify the first particle-particle pair propagator by the first 
and third row entries. The propagator is  $\Pit^{(2)}_{\gvec{\sigma}, \vec{s}}$, where $\gvec{\sigma}=(\sigma, -\sigma)$, because the first and third Green's functions
have spin $\sigma$ and $-\sigma$ respectively, and $\vec{s}=(1,1)$ because $\gvec{f}_{11} = \gvec{f}_{31} = 1$.
Having `used' the Green's functions in question, we mask their corresponding entries in $\gvec{f}$ by appending `1' to $L_c$ and `1,3' to $L_r$ so that $L_c = \{1\}$ and $L_r = \{1,3\}$.  We temporarily associate this
propagator with the matrices
\begin{equation}
\vec{q}_1 = 
\begin{bmatrix}
1 & 0 & -1 \\
1 & 0 &  0 \\
\end{bmatrix},\quad
\gvec{\Omega}_{\vec{q}_1} = 
\begin{bmatrix}
\Omega \\ 0 
\end{bmatrix}
,
\label{eq:q1}
\end{equation} 
constructed from the first and third rows of $\gvec{f}$ and $\gvec{\Omega}^{(p)}$ in Eq.~\eqref{eq:fmatrixeq}. We now examine the second
column of $\gvec{f}$ in Eq.~\eqref{eq:fmatrixeq} and note that the two non-zero entries are contained in the second and fourth rows, neither of which is in $L_r$. This is thus another
instance of $\Pi^{(2)}_{(\sigma, -\sigma); (1,1)}$, and after updating $L_c = \{1,2\}$, and $L_r = \{1,3,2,4\}$. We associate this propagator with a $\vec{q}_2$ matrix
equal to Eq.~\eqref{eq:q1}. We proceed to the third column and examine the possibility that it corresponds to a triple propagator. We conclude 
that it does not,  since its non-zero elements appear on rows $1,2,5$, the first two of which are already in $L_r$. Having exhausted 
the columns of $\gvec{f}$, the final step is to remove all columns in $L_c$ and all rows in $L_r$ from $\gvec{f}$ to obtain $\vec{g}$, and all 
column in $L_c$ from $\vec{q}_1$, $\vec{q}_2$ to obtain their final simplified form and remove the references to the variables of integration that have been 
eliminated. To conclude the procedure we also delete all rows in $L_r$ from $\gvec{\Omega}^{(p)}$, to obtain $\gvec{\Omega}^{(g)}$.
We thus have $\gvec{f} = [1]$, $\gvec{\Omega}^{(g)} = [0]$  and
\begin{equation}
\vec{q}_1 = \vec{q}_2 = 
\begin{bmatrix}
1 \\
0 \\
\end{bmatrix}, 
\qquad \gvec{\Omega}_{\vec{q}_2} = \gvec{\Omega}_{\vec{q}_1} = 
\begin{bmatrix}
\Omega \\ 0 
\end{bmatrix},
\end{equation}
and the amplitude in Eq.~\eqref{eq:exampleeq} simplifies to
\begin{equation}
\int \rmd \epsilon_1 \GeeNt_\sigma(\epsilon_1)\left[\Pi^{(2)}_{(\sigma, -\sigma); (1,1)}(\epsilon_1 + \Omega,0)\right]^2.
\end{equation}

The final step in our calculation is the numerical integration, which is handled with adaptive quadrature~\cite{GSL} and cubature methods~\cite{sjohnson}.
By evaluating \emph{vectors} of points at every iteration, the calculation can be seamlessly parallelized to 
take advantage of modern computer architectures. 
Note that it is possible, especially in the absence of a magnetic field, for two distinct diagrams to be equal numerically. We test for this by
evaluating each diagram individually at a randomly chosen frequency of order $\Deltat$, and, if an equal pair is found, ensuring that only one of the diagrams is included 
in the integration, with an adjusted prefactor, is included in the integration. A
similar situation occurs in the case of particle-hole symmetry, where $\tilde{\epsilon_d} = 0$, rendering $\GeeNt(\omega)$ an odd function of 
frequency and leading to the numerical cancellation of several diagrams.  

\subsection{Assembling the RPT}

So far we have been working order-by-order in the effective interaction $\Ueff = \Ut + \lambda_3$, expressing the self-energy as 
\begin{equation}
\Sigmat(\omega) = \sum_{n=2}  \gamma_n (\omega) \Ueff^n.
\label{eq:sigmatseries}
\end{equation} 
Ultimately, we aim to obtain $\Sigmat(\omega)$ as a power series in $\Ut$ rather than $\Ueff$. To accomplish this, we note
that the counter-term $\lambda_3$ is defined as in Eq.~\eqref{eq:RGeq}. By calculating the renormalized four-vertex at zero frequency
we can thus obtain $\Ut$ as a power series in $\Ueff$, 
\begin{equation}
\Ut = \Gammat_{\uparrow \downarrow}(0,0;0,0) = \Ueff + \sum_{n=2}^\infty \alpha_n \Ueff^n.
\label{eq:utseries}
\end{equation}
Working order-by-order we can invert this equation:
\begin{equation}
\Ueff = \Ut + \sum_{n=2}^\infty \beta_n \Ut^n, 
\label{eq:ueffseries}
\end{equation}
where 
\begin{align}
\beta_2 &= -\alpha_2, \nonumber \\
\beta_3 &= 2 \alpha_2^2-\alpha_3,\nonumber \\
\beta_4 &= -5 \alpha_2^3+5 \alpha_2 \alpha_3-\alpha_4,\nonumber \\
\beta_5 &= 14 \alpha_2^4-21 \alpha_2^2 \alpha_3+6 \alpha_2 \alpha_4+3 \alpha_3^2-\alpha_5.
\end{align}
Thus, the calculation of the self-energy in terms of $\Ueff$ can be rewritten as a series in $\Ut$
\begin{equation}
\Sigmat(\omega) = \sum_{n=2}^\infty \delta_n(\omega)  \Ut^n,
\end{equation}
where  
\begin{align}
\delta_2 &= \gamma_2 (\omega),\nonumber \\
\delta_3 &= -2 \alpha_2 \gamma_2(\omega)+\gamma_3(\omega), \nonumber \\
\delta_4 &= \left(5 \alpha_2^2-2 \alpha_3\right) \gamma_2(\omega)-3 \alpha_2 \gamma_3(\omega)+\gamma_4(\omega), \nonumber \\
\delta_5 &= \left( -14 \alpha_2^3 +12 \alpha_3\alpha_2 - 2\alpha_4 \right)\gamma_2(\omega),    \nonumber \\
 &\qquad            + \left(9\alpha_2^2- 3\alpha_3\right)\gamma_3(\omega)      - 4 \alpha_2 \gamma_4(\omega)        + \gamma_5(\omega).
\end{align}
These relations enable us to deduce the renormalized expansion from the bare one and show explicitly how the inclusion of
counter-terms results in the re-organization of the series.

\subsection{Checks}

To check our calculation for the self-energy we can relate the RPT to the perturbation theory of Yamada and Yosida~\cite{Yosida70, Yamada75, Yosida75} by replacing $\Ueff$ in Eq.~\eqref{eq:sigmatseries}
with the bare $U$, the parameters $\tilde{\epsilon_d}, \Deltat$ with their bare counter-parts and setting all the counter-terms equal to zero. 
Similarly, we can check the four-vertex against the calculation of Ref.~\cite{Hewson01}. We find that the analytic results 
\begin{align}   
&\partial_\omega \Sigma(0) = -\left(3-\frac{\pi^2}{4}\right) u^2 - \left(105 - \frac{45\pi^2}{4} + \frac{\pi^4}{16}\right)u^4 + \ldots \nonumber\\
&\Gamma_{\uparrow \downarrow}(0,0;0,0) = U \left[ 1 + \left(15-\frac{3\pi^2}{2}\right)u^2 + \ldots \right],
\end{align}

where $u=U/\pi\Delta$, are reproduced by our calculation.

\section{Results}

In this section we present numerical results for the irreducible self-energy and resultant spectral density. For all the calculations we fix $\pi\Delta = D/100$, 
where $D$ is the conduction band width, and $U=3\pi\Delta$. In our discussion we consider the following parameter configurations: 
\begin{inparaenum}[(i)]
\item  a symmetric model with $\epsilon_d=-U/2$;
\item  a model with weak asymmetry $\epsilon_d=-1.2\pi\Delta$; and
\item  a model with pronounced asymmetry and $\epsilon_d = -3\pi\Delta$.
\end{inparaenum}
In all cases, the scale of the problem is set by the renormalized density of states at the Fermi level, 
\begin{equation}
\tilde{\rho_0} = \frac{\Deltat/\pi}{\tilde{\epsilon_d}^2 + \Deltat^2}.
\label{eq:rho0}
\end{equation}
We use $\tilde{\rho_0}$ to define the Kondo temperature $T_K$ as $T_K =  1/4\tilde{\rho_0}$; in the Kondo limit this reduces
to the usual definition of $T_K$ in terms of the susceptibility~\cite{Hewson01}. We will 

\subsection{Self-energy}

For comparison purposes we will juxtapose the renormalized self-energy obtained from RPT
with the corresponding result obtained from the NRG. Since NRG calculations are set up
with the bare self-energy in mind, we have to use Eq.~\eqref{eq:sigmarelation} to relate
the two quantities. For the imaginary part we find that 
\begin{equation}
\Im\Sigmat_\sigma(\omega) = z_\sigma \Im\Sigma_\sigma(\omega).
\label{eq:imsigmats}
\end{equation}
We remark that inaccuracies in our NRG calculation result in a slightly non-zero $\Im\Sigmat_\sigma(0)$;
to correct for this we offset our results by a small imaginary number. An equation similar to Eq.~\eqref{eq:imsigmats} can be 
derived for the real part; in practice however, it is of limited use.
The reason is that the renormalized parameters are not determined from the NRG using the definitions in Eq.~\eqref{eq:renormdef}
but from the effective linear chain Hamiltonian (for more information see the Appendix of Ref.~\cite{Hewson04}).
Since the low-energy limit of the real part of Eq.~\eqref{eq:sigmarelation} relies crucially on the numerical cancellation
between $\Sigma(\omega)$ and $(z-1)\omega$, we found it difficult to obtain a result for $\Re\Sigmat(\omega)$ with a 
vanishing derivative at zero frequency. It is for this reason that we only show NRG results for $\Im\Sigmat(\omega)$.
Note that due to the NRG's successive elimination of higher energy scales we expect that its estimate for $\Sigmat(\omega)$ will
only be accurate in the low frequency sector. 

\begin{figure*} 
\centering
\begin{subfigure}[b]{0.49\linewidth} 
\includegraphics[scale=0.31]{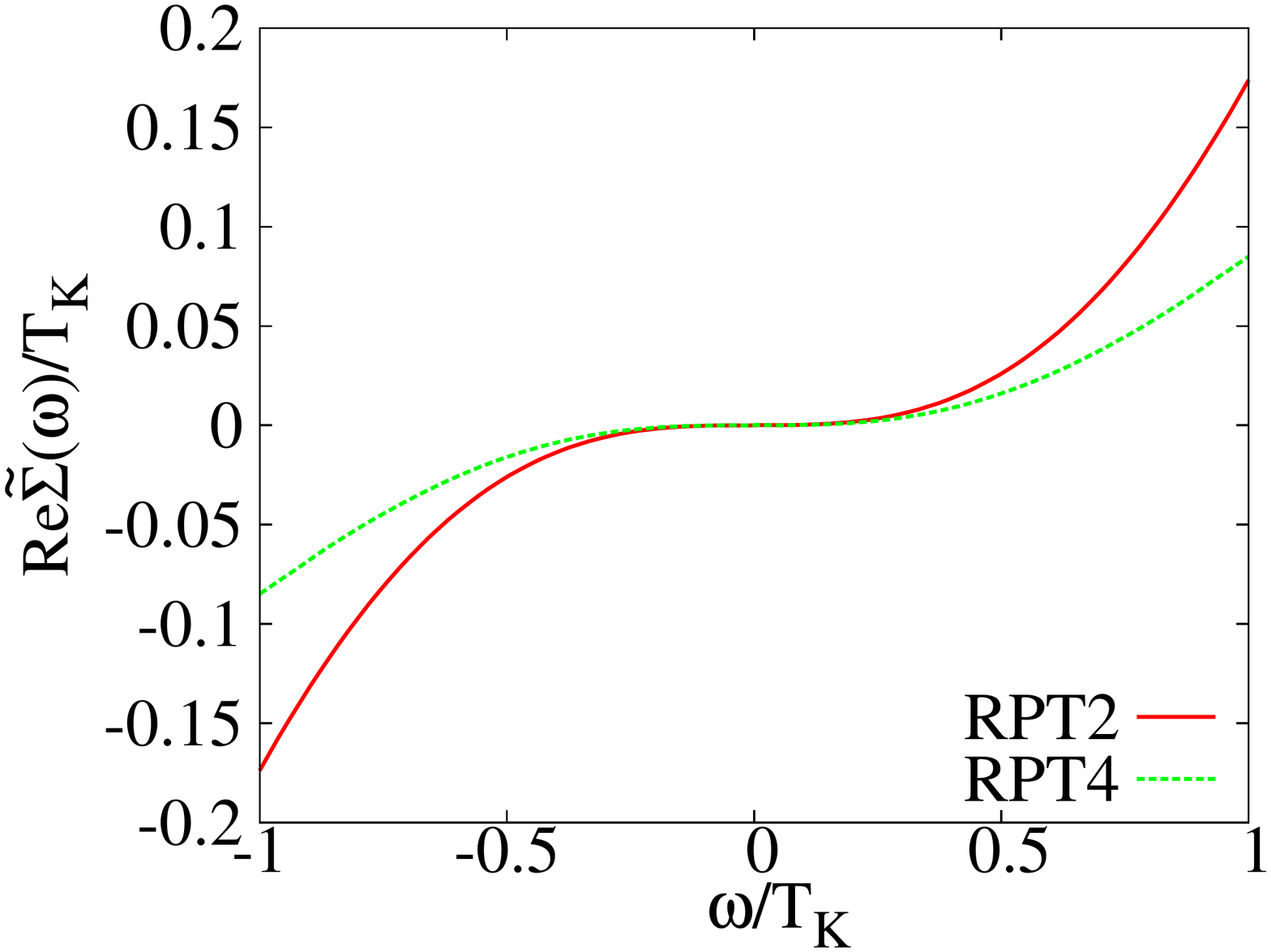}
\end{subfigure} 
\begin{subfigure}[b]{0.49\linewidth} 
\includegraphics[scale=0.31]{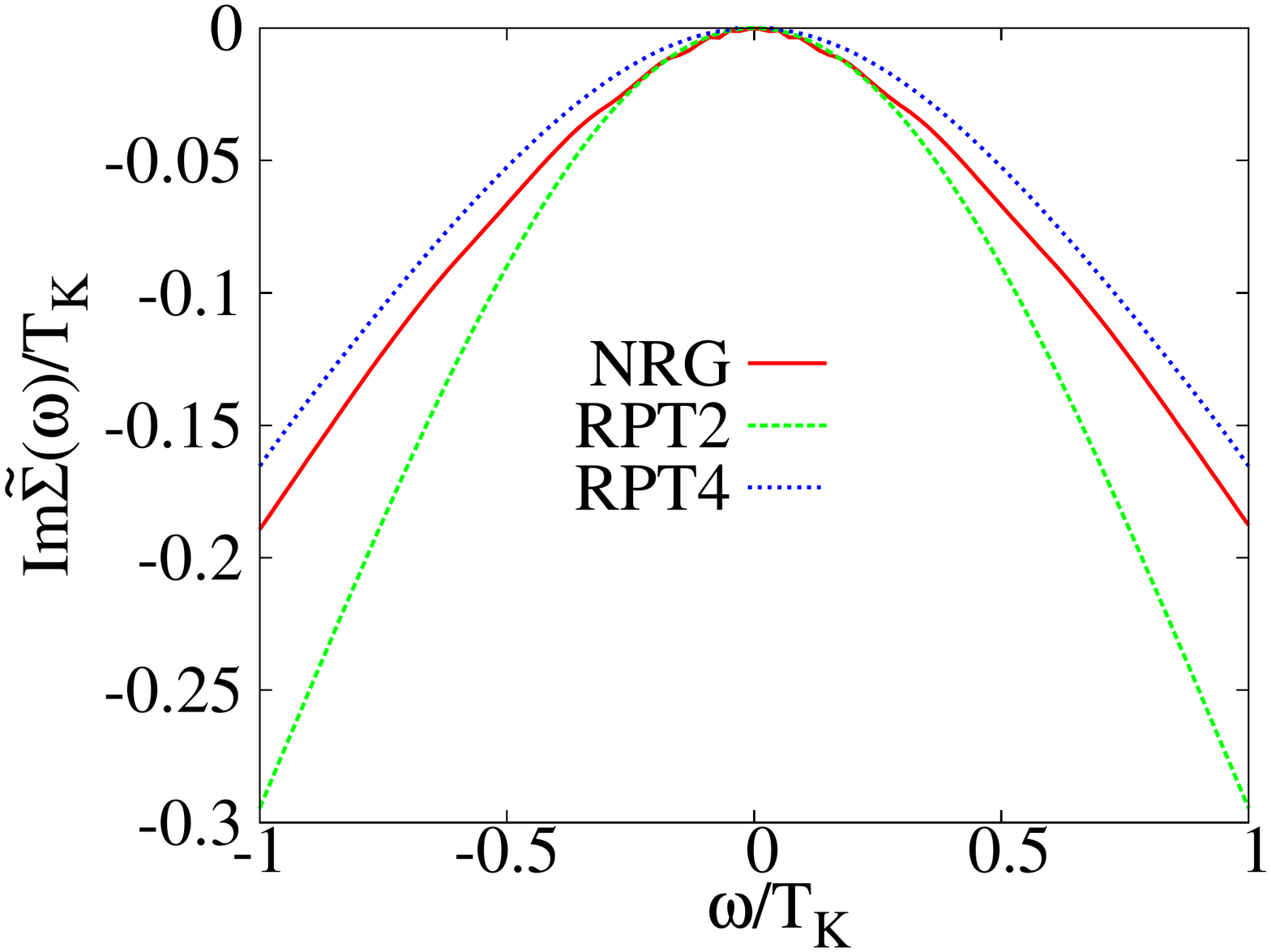}
\end{subfigure}
\caption{The real (left) and imaginary (right) parts of the renormalized self-energy for the particle-hole symmetric model.  }
\label{fig:sigmaph}
\end{figure*}

\begin{figure*} 
\centering
\begin{subfigure}[b]{0.49\linewidth} 
\includegraphics[scale=0.31]{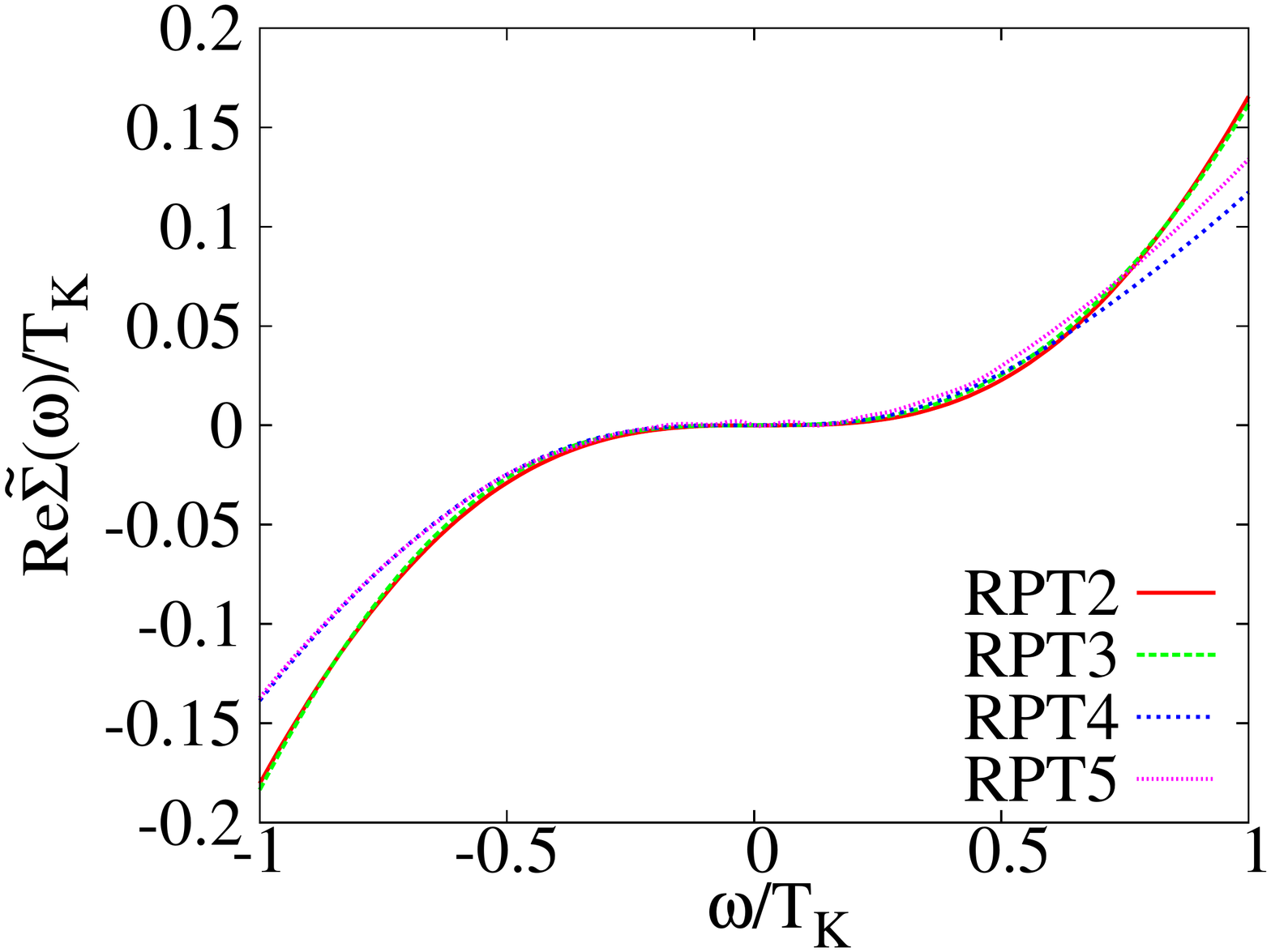}
\end{subfigure} 
\begin{subfigure}[b]{0.49\linewidth} 
\includegraphics[scale=0.31]{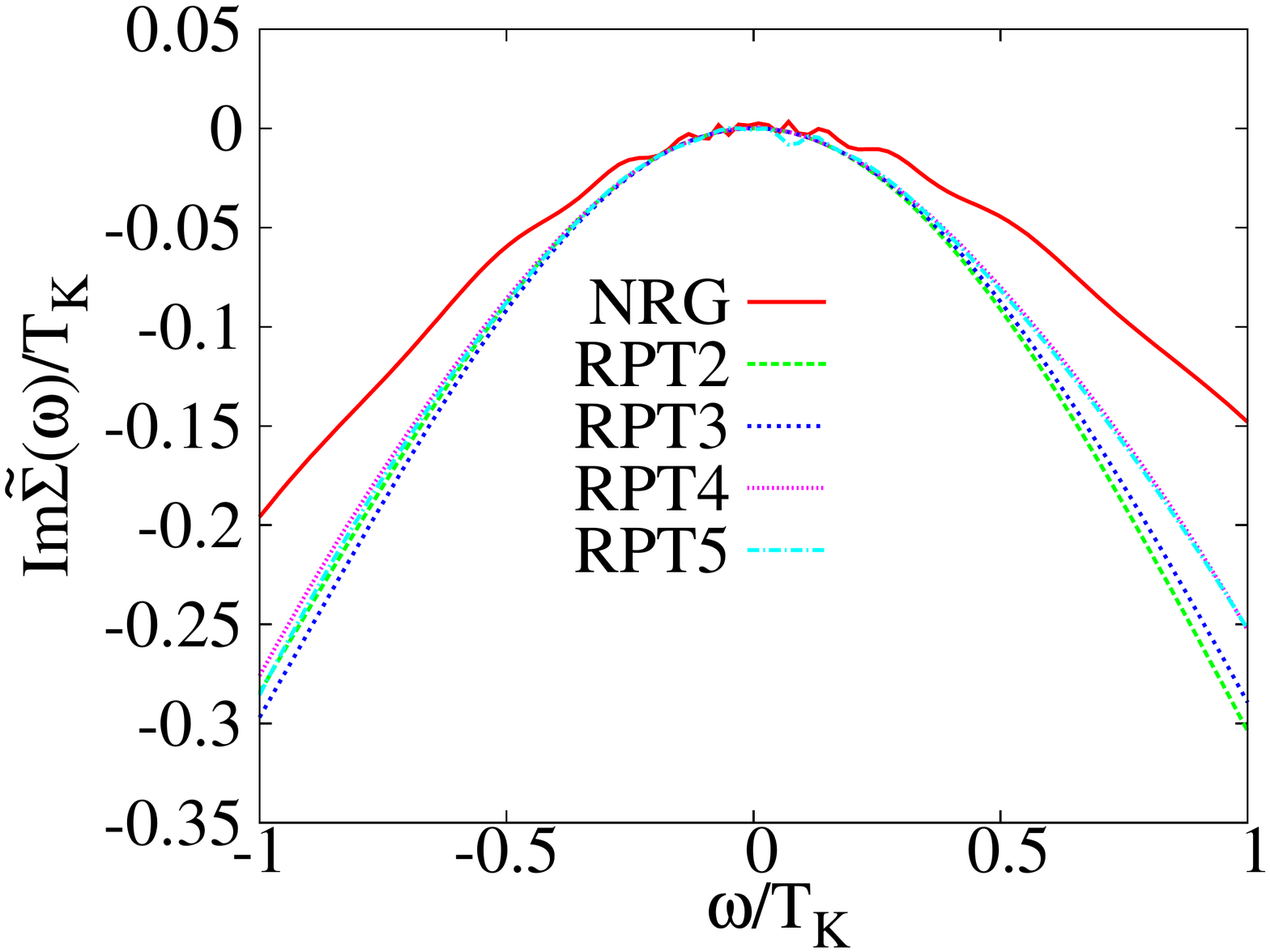}
\end{subfigure}
\caption{The real (left) and imaginary (right) parts of the renormalized self-energy for the weakly asymmetric model (model (ii)).}
\label{fig:sigmaweak}
\end{figure*}

\begin{figure*} 
\centering
\begin{subfigure}[b]{0.49\linewidth} 
\includegraphics[scale=0.31]{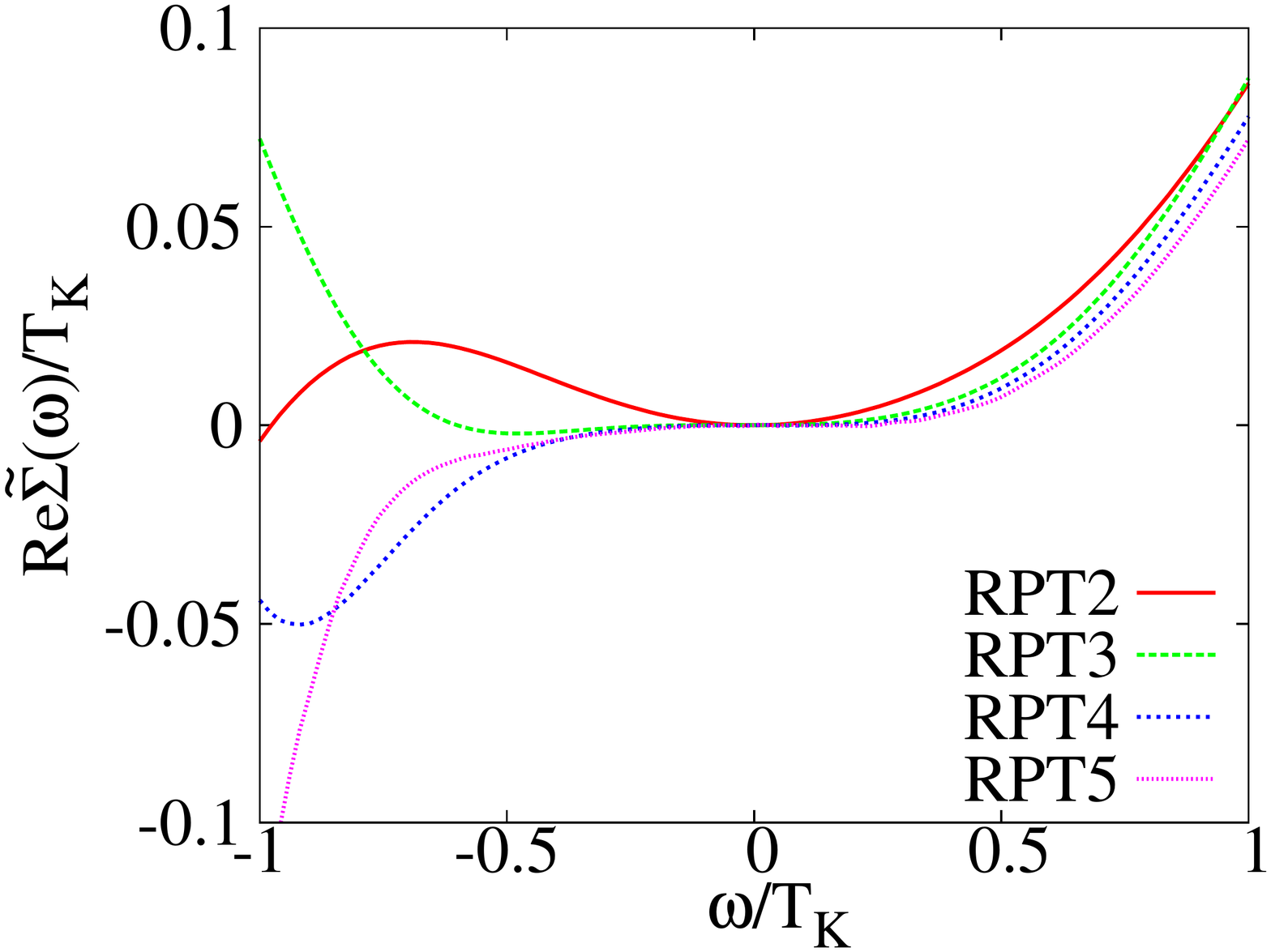}
\end{subfigure} 
\begin{subfigure}[b]{0.49\linewidth} 
\includegraphics[scale=0.31]{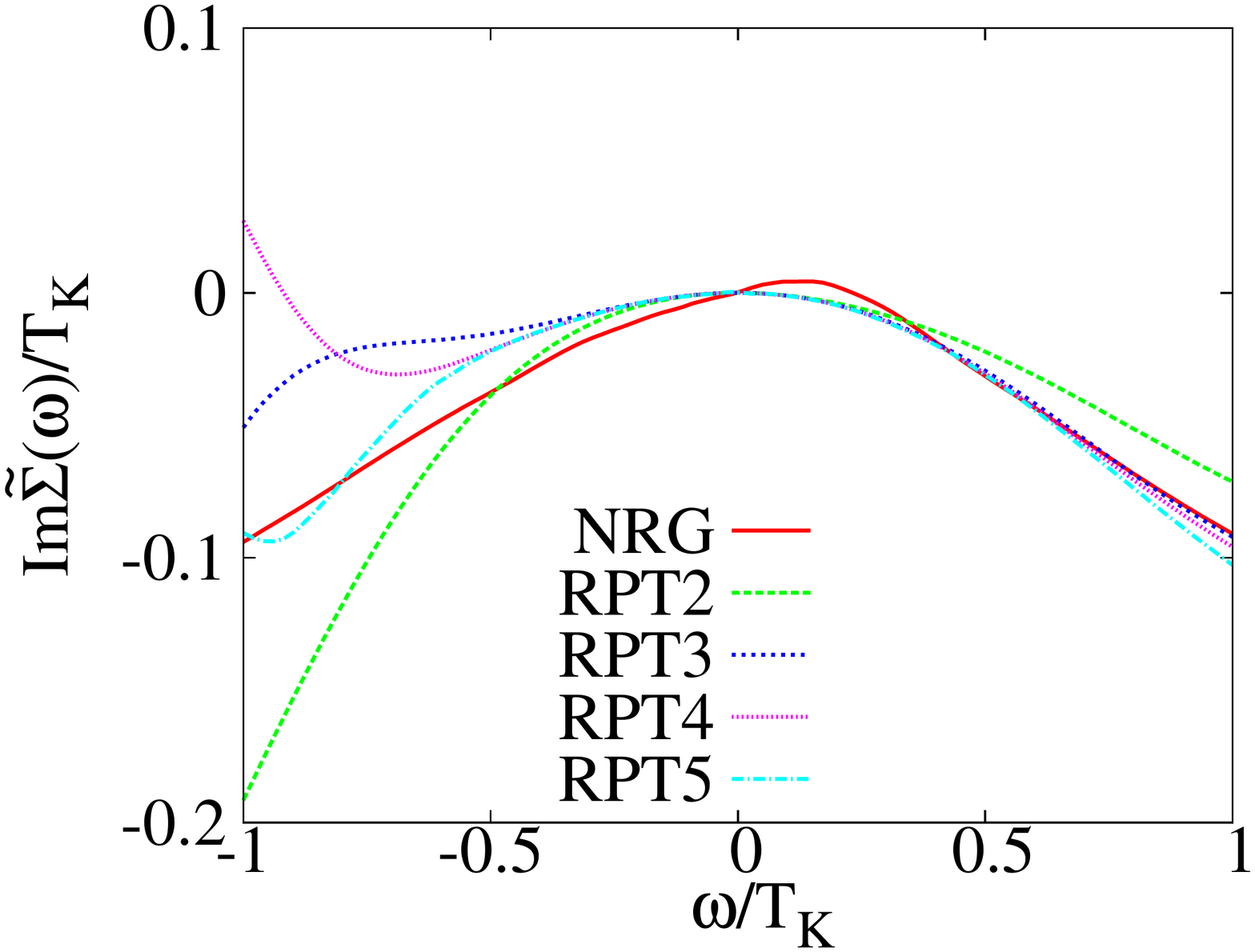}
\end{subfigure}
\caption{The real (left) and imaginary (right) parts of the renormalized self-energy for the strongly asymmetric model (model (iii)). }
\label{fig:sigmastrong}
\end{figure*}

We begin by discussing the symmetric model (case (i)) . As we have already remarked, in the presence of particle-hole symmetry $\tilde{\epsilon_d}=0$ and the propagator is an odd
function of frequency. The parity of the Green's function also results in the cancellation of the particle-hole and particle-particle propagators
\begin{equation}
\Pit^{(2)}_{\gvec{\sigma}; (1, 1)}(\omega_1, \omega_2) = -\Pit^{(2)}_{\gvec{\sigma}; (1, -1)}(\omega_1, \omega_2), 
\end{equation}
and consequently the odd-order terms of the self-energy vanish for all frequencies. Using the method described in Ref.~\cite{Hewson04} we find that $\Deltat = 2.54\times10^{-4}$ and $\Ut = 7.95\times10^{-4}$,
giving $\Ut/\pi\Deltat = 0.994$. That this ratio is almost 1 is no coincidence; in the Kondo limit a universal scale, the Kondo temperature $T_K$ emerges and $\Ut \rightarrow 4T_K, \pi\Deltat \rightarrow 4T_K$~\cite{Hewson93b}.

In all cases,  the RPT estimate of the renormalized self-energy will always be exact in the limit $\omega\rightarrow0$, in the sense of Eq.~\eqref{eq:RGeq}. 
At particle-hole symmetry, and for small but finite frequencies $|\omega \tilde{\rho_0}| \ll 1$, the $\omega^2$ coefficient of $\Sigmat(\omega)$ is reproduced 
exactly~\cite{Hewson93b} by the second-order calculation; the fourth order term does not contribute terms of order $\omega^2$. As we increase $|\omega|$ 
the dominant term is the $\omega^4$ term, which is not exactly given by the second-order calculation; this is corrected by the fourth-order contribution, extending 
the domain of validity of the RPT. At higher frequencies the disparity between the two RPT curves suggests the breakdown of the expansion; to obtain reliable
results one must calculate the higher-order terms. We thus expect more elaborate approximations within RPT to continually extend the domain of validity of the resultant $\Sigmat(\omega)$.
Were we to have, somehow, the ability to take into account \emph{all} Feynman diagrams to all orders we would recover a $\Sigmat(\omega)$ exact for all frequencies, though
clearly then we would also be able to calculate $\Sigma(\omega)$ in the first place. 

Next, we discuss a slightly asymmetric model with $\epsilon_d = -1.2\pi\Delta$, for which we find that $\tilde{\epsilon_d} = 2.14\times10^{-5}$, $\Deltat=2.93\times10^{-4}$ and
$\Ut = 9.18\times10^{-4}$. The non-zero but small in magnitude $\tilde{\epsilon_d}$ now gives rise to odd-order terms in $\Sigmat(\omega)$. From Fig.~\eqref{fig:sigmaweak} we
see that the third and fifth order terms are small in value and essentially only slightly modify the second and fourth order curves respectively. To determine the stability
of the series to order $n$ we see that to simply look at the $n+1$ term is not sufficient, since the $n+2$ terms can still remain important. 

Finally, we turn our attention to Fig.~\eqref{fig:sigmastrong} and case (iii), a  very asymmetric model with $\epsilon_d = - 3\pi\Delta$. For small $|\omega|$ we find
again that the RPT is in good agreement with the NRG and that the lower order contributions dominate the result. A dramatic breakdown of the expansion, in both the real
and imaginary components,  is evident for small negative value of $\omega$ in contrast to cases (i) and (ii) where it is reasonably well-behaved even around $\omega\tilde{\rho}_0 \approx 0.5$. 

\subsection{Spectral densities}

\begin{figure} 
\includegraphics[scale=0.28]{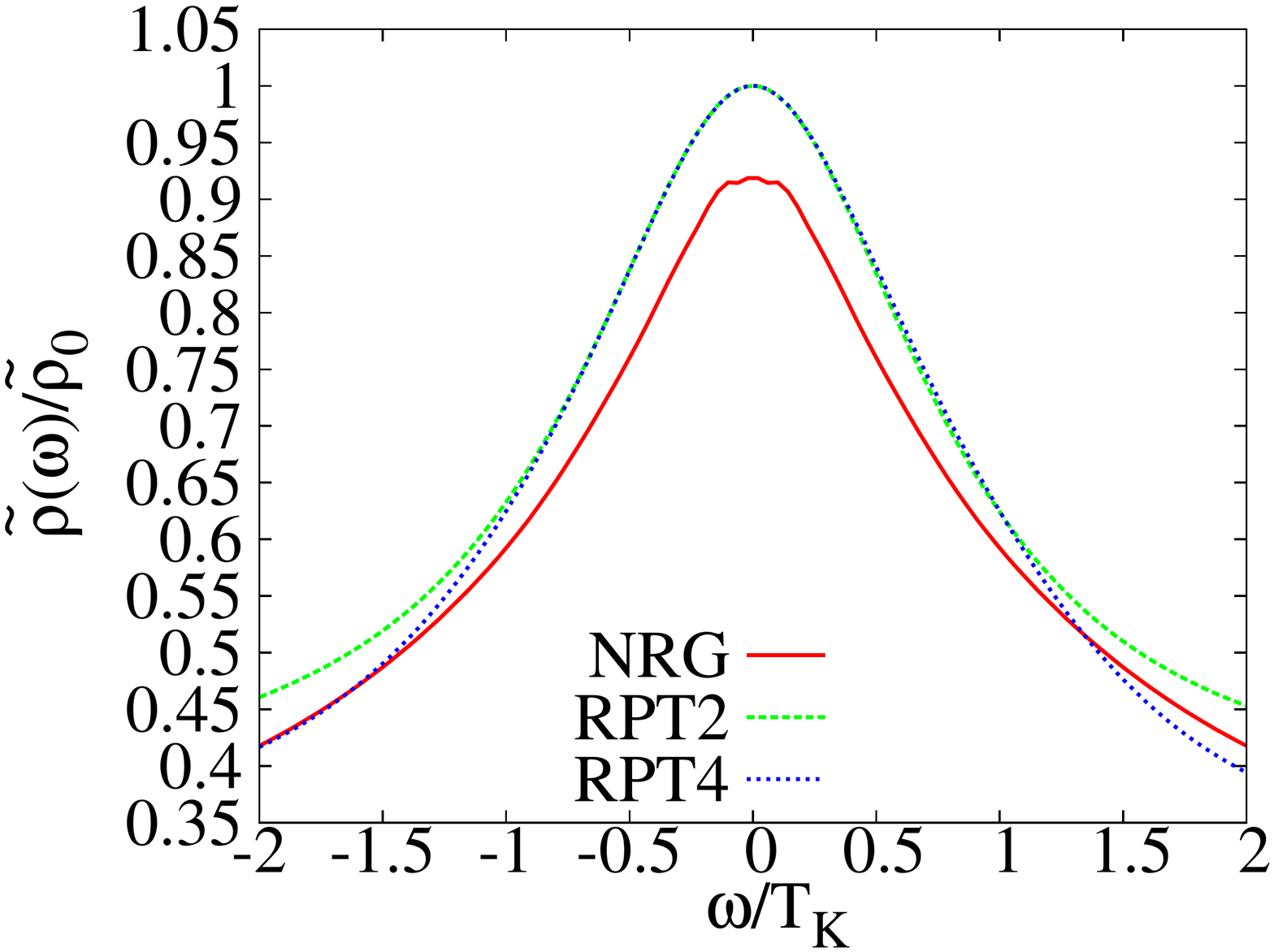}
\caption{The spectral density for the particle-hole symmetric model (model (i)).}
\label{fig:specph}
\end{figure}

\begin{figure} 
\includegraphics[scale=0.28]{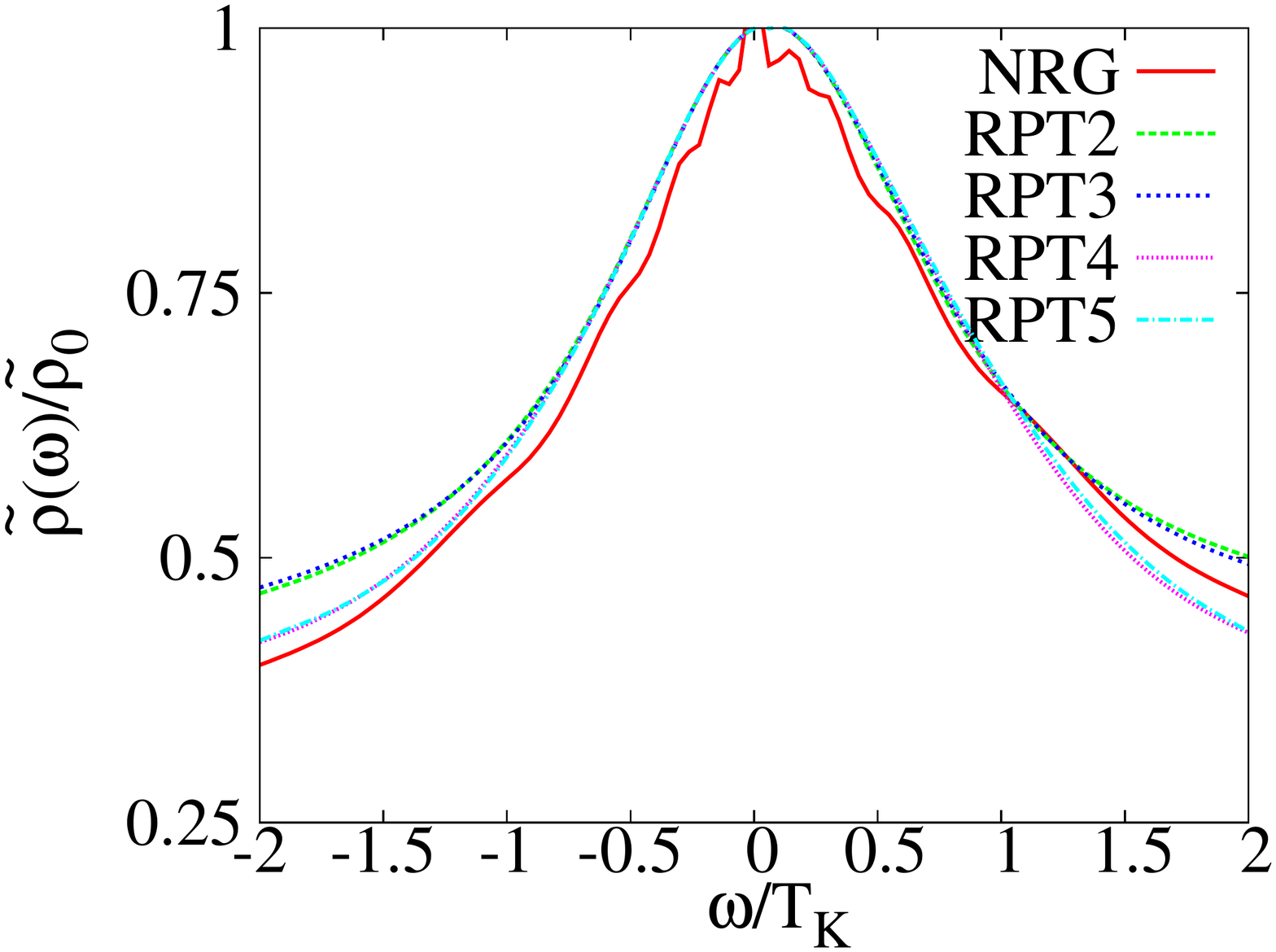}
\caption{The spectral density for model (ii), with some asymmetry.}
\label{fig:specweak}
\end{figure}

\begin{figure} 
\includegraphics[scale=0.28]{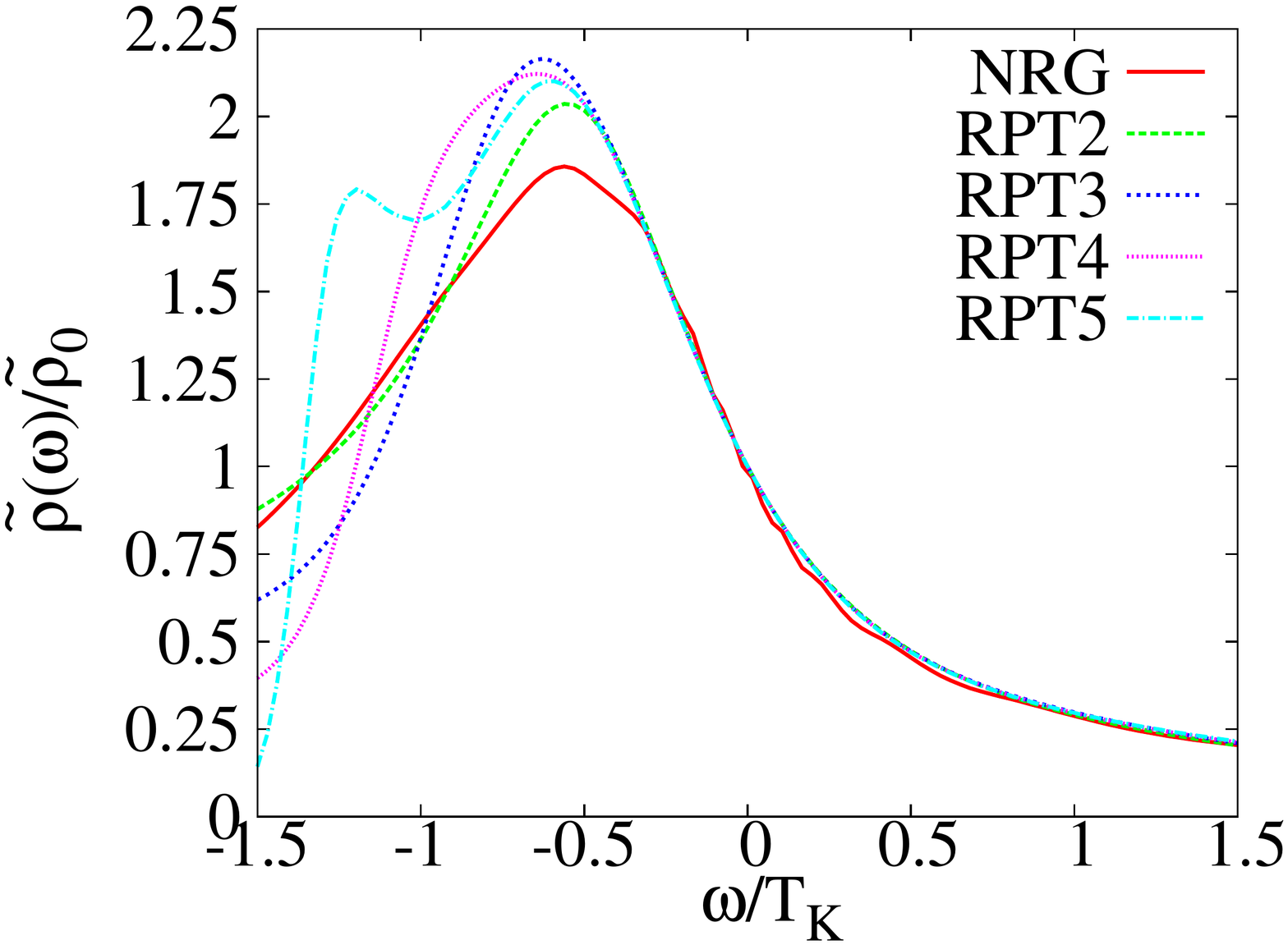}
\caption{The spectral density for the very asymmetric model (model (iii)).}
\label{fig:specstrong}
\end{figure}

We define the interacting Green's function 
\begin{equation}
\tilde{G}_\sigma (\omega) = \frac{1}{\omega - \tilde{\epsilon_d} + i\Deltat - \Sigmat^R(\omega)},
\end{equation}
where $\Sigmat^R(\omega)$ is the \emph{retarded} self-energy, which, as usual, differs from the
causal $\Sigmat(\omega)$ one obtains from the diagrammatics only by a $\sgn(\omega)$ term in the imaginary part.
We now turn our attention to the quasi-particle spectral density, defined as $\tilde{\rho}(\omega) = -\frac{1}{\pi}\Im\tilde{G}(\omega)$
and express it in terms of the renormalized parameters as 
\begin{equation}
\tilde{\rho}(\omega) = \frac{1}{\pi} \frac{\Deltat - \Im\Sigmat^R(\omega)}{(\omega - \tilde{\epsilon_d} - \Re\Sigmat^R(\omega))^2 + (\Deltat - \Im\Sigmat^R(\omega))^2}.
\label{eq:rhot}
\end{equation}
Note that the spectral density is sensitive to the $\emph{reducible}$ self-energy. The low-frequency properties of the $n$'th order
spectral density are thus the result of competition between higher-order irreducible self-energies and powers of $\GeeNt(\omega)$ combined 
with powers of lower-order irreducible self-energies. 

Results for cases (i), (ii) and (iii) are shown in Figs.~\ref{fig:specph},~\ref{fig:specweak} and \ref{fig:specstrong} respectively.
To avoid the instability in the real part of $\Sigmat(\omega)$ we do not use Eq.~\eqref{eq:rhot} directly to extract the result from the NRG;
instead we use the NRG result for $\rho(\omega)$ and, from Eq.~\eqref{eq:grpt}, $\tilde{\rho}(\omega) = \rho(\omega)/z$. Note that
$\tilde{\rho}_0$ is calculated from the renormalised parameters, and consequently, due to inaccuracies in our NRG calculation, the ratio
$\rho(\omega=0)/\tilde{\rho}_0$ may differ from $1$.

For the particle-hole symmetric model of case (i) we find again that the second and fourth order terms coincide for small $\omega$
and start deviating from each other and the NRG result as $|\omega|$ is increased, with the RPT4 curve being in closer agreement
with the latter. This picture persists in Fig.~\eqref{fig:specweak} where, as in the case of the self-energy,  two pairs of similar curves 
emerge.  We remark that at particle-hole symmetry $\tilde{\rho}(\omega=0)=(\pi\Deltat)^{-1}$; away from particle-hole symmetry $\tilde{\rho}(\omega=0)$ is
given exactly by the renormalized parameters as per Eq.~\eqref{eq:rho0}.

\section{Conclusion}

In this paper we have presented a relatively simple way of automating the calculation of the renormalized self-energy so
that it can be carried out by a computer without user intervention. Our presentation was logically partitioned into
three steps: the diagram generation, the application of the rules and analytic simplification of the integrals and 
the numerical integration of the diagrams.

To illustrate the usefulness of the method we have calculated the self-energy
up to fifth-order inclusive, a calculation which would otherwise be extremely tedious to perform by hand. We performed
the calculation for three possible values of the asymmetry and found the results to be in good agreement with the NRG
in the meaningful low frequency region. In all cases we find that the higher order term's contribute more at higher
frequencies, with RPT becoming increasingly more accurate as $\omega \rightarrow 0$. 

Though our discussion has been confined to the  single-impurity Anderson model, the method here is readily generalisable to other models by 
replacing the Green's function with the appropriate one. Unfortunately, the calculation of the $k$-particle/hole propagator
of the Appendix relies on the linear dependence of $[\GeeNt(\omega)]^{-1}$ on $\omega$ and thus does not generalize. This
is not too serious an obstacle, for we can always numerically evaluate and tabulate the cases $k=2,3$ which
most commonly appear. A further generalization can be realized by replacing the scalar propagator with a matrix 
quantity to obtain calculate the behaviour of the impurity out of equilibrium~\cite{Oguri01}. This is of particular
importance in the study of quantum dots under a bias voltage. 

\begin{acknowledgments}
V.P.\ would like to acknowledge the financial support of the Engineering and Physical Sciences Research Council. 
\end{acknowledgments}

\FloatBarrier
\appendix*
\section{Appendix}

In Eq.~\eqref{eq:npp} we define a $k$-particle/hole propagator as 
\begin{equation}
\Pit^{(k)}_{\vec{\sigma}; \vec{s}}(\omega_1, \ldots, \omega_n) = i^n\int_{-\infty}^\infty \rmd  \omega' \prod_{i=1}^{k} G^{[0]}_{\sigma_i} (s_i\omega' + \omega_i).
\label{eq:npp2}
\end{equation}
where  $s_i \in \{-1,1\}$. Due to the $\sgn$ term in the causal Green's function (Eq.~\eqref{eq:greens}, the integrand can be thought of as a piecewise function in
$\omega'$. We begin by identifying the points $y_i$, which we will call \emph{nodes}, where $s_iy_i + \omega_i = 0$.
We can ensure by appropriate labelling of the nodes that $ -\infty \geq y_1 \geq y_2 \ldots \geq y_n$. For brevity, we write the $i$'th Green's function in the form
\begin{equation}
\GeeNt_{\sigma_i} (\omega) = \frac{s_i}{\omega - \alpha_i(\omega)},
\label{eq:geentpf}
\end{equation}
where $\alpha_i = s_i\left[ -\omega_i + \epsilon_{d, \sigma_i} - i\Delta\sgn(s_i\omega+\omega_i)\right]$. We temporarily make the assumption, which will be
lifted later, that all the $\alpha_i$ are distinct. Written as a product of terms of the form of Eq.~\eqref{eq:geentpf}, the integrand depends explicitly on $\omega'$ but
also implicitly through the dependence of the $\alpha$. Having identified the nodes we can rewrite Eq.~\eqref{eq:npp2} as
\begin{equation}
i^{-n}\Pit^{(n)}_{\vec{\sigma}; \vec{s}}(\omega_1, \ldots, \omega_n) = J(y_1, y_n) + \sum_{i=1}^{n-1} F(y_i, y_{i+1}), 
\end{equation}
where
\begin{align}
F(y_i, y_{i+1}) &= \int_{y_i}^{y_{i+1}} \rmd\omega'  \prod_{i=1}^{k} G^{[0]}_{\sigma_i} (s_i\omega' + \omega_i) \nonumber \\
J(y_1, y_n) &=  \lim_{\Lambda\rightarrow\infty} \Bigg\{ \int_{-\Lambda}^{y_1} \rmd\omega'  \prod_{i=1}^{k} G^{[0]}_{\sigma_i} (s_i\omega' + \omega_i)\\ 
&  \qquad +  \int_{y_n}^{\Lambda} \rmd\omega'\prod_{i=1}^{k} G^{[0]}_{\sigma_i} (s_i\omega' + \omega_i) \Bigg\}.
\end{align}
The decomposition of the real axis into intervals on which the integrand does not change form means that we can perform a partial fraction
decomposition with coefficients specific to the region. Hence we can write
\begin{equation} 
\frac{1}{(\omega' - \alpha_1)(\omega' - \alpha_2)\ldots(\omega'-\alpha_n)} = \sum_{i=1}^{n} \frac{\beta_i}{\omega' - \alpha_i},
\end{equation}
where $\beta_j = 1/f'(\alpha_j)$, $f(\omega) = (\omega-\alpha_1)\ldots(\omega-\alpha_n)$. We can now simply integrate each partial fraction
separately. Let   $\vec{\alpha}^{(i)}$ and $\vec{\beta}^{(i)}$ denote the values of the relevant quantities in the region $(y_i, y_{i+1})$. Then
\begin{align}
F(y_i, y_{i+1}) &= \sum_{j=1}^n \beta^{(i)}_j \mathrm{Ln} \left(\frac{y_{i+1} - \alpha^{(i)}_j}{y_i - \alpha^{(i)}_j}\right) \nonumber \\
J(y_1, y_n) &= \sum_{j=1}^n \Big[\beta^{(0)}_j \mathrm{Ln} \left(y_{1} - \alpha^{(0)}_j\right) -    \nonumber \\
& \qquad \beta^{(n+1)}_j  \mathrm{Ln} \left({y_n - \alpha^{(n+1)}_j}\right)\Big], 
\end{align}
where $\mathrm{Ln}(z)$ denotes the principal branch of the complex logarithm defined as $\ln|z| + i\Arg(z)$, $-\pi<\Arg(z)<\pi$  and
the labels $0, n+1$ denote the values of the underlying quantities in the intervals $(-\infty, y_1)$ and $(y_n , \infty)$ respectively.

In practical applications we may encounter numerical difficulties if the $\omega_i$ are such that
any two $\alpha_i$ in a particular region coincide, or nearly coincide This will cause our partial fraction to break down.
 We deal with this in a crude yet effective manner: when any $\alpha_i$, $\alpha_j$ are
too close to each other, we separate them by a very small, arbitrary constant. After separating the offending $\alpha_i, \alpha_j$ it is
important to update the values of the corresponding $y_i, y_j$ to ensure the consistency of the calculation.

\bibliographystyle{apsrev4-1}
\bibliography{bibliography}
\end{document}